# Collective Memory in the Digital Age

Taha Yasseri[1,2]*, Patrick Gildersleve[3], and Lea David[1]

[1] School of Sociology, University College Dublin, Dublin, Ireland
[2] Geary Institute for Public Policy, University College Dublin, Dublin, Ireland
[3] Department of Methodology, London School of Economics and Political Science, London, UK

*Corresponding Author: taha.yasseri@ucd.ie

**Abstract**
The digital transformation of our societies and in particular information and communication technologies have revolutionized how we generate, communicate, and acquire information. Collective memory as a core and unifying force in our societies has not been an exception among many societal concepts which have been revolutionized through digital transformation. In this chapter, we have distinguished between "the digitalized collective memory" and "collective memory in the digital age". In addition to discussing these two main concepts, we discuss how digital tools and trace data can open doorways into the study of collective memory that is formed inside and outside of the digital space.

***Keywords***: Collective memory; Digital Age; Web 2.0; Digitalized Memory; Digital Memory

"If we want to bring the memory of the Holocaust to the young generation, we have to bring it to where they are," said the project co-producer, Mati Kochavi, an Israeli high-tech billionaire who hails from a family of Holocaust victims, survivors and educators. "And they're on Instagram."[1]

# Introduction

Our societies form, sustain, and function because of our intelligence, the ability to learn and memorize. In the animal kingdom, the smarter animals are, the more social they are. Dunbar's social brain theory (Dunbar, 2006) even suggests that significantly superior human intelligence compared with other primates is the result of our need to be able to manage and maintain our social lives; it is humans' outstanding sociality and attraction to collectivities that forced them to become smart. The bottom line is, that in human societies, collectives are intelligent and intelligence is collective.

---

[1] https://www.timesofisrael.com/instagram-story-of-young-holocaust-victim-aims-at-new-generation/

Whilst memory is only one form of intelligence, it is a crucial element separating human societies from other forms of intelligent groups. In the social context, the memory always reflects present needs (Nora, 1989; Connerton, 1989; Lowenthal, 1985; Zerubavel, 1995; Kuljić, 2006), and "while the object of commemoration is usually to be found in the past, the issue which motivates its selection and shaping is always to be found among the concerns of the present" (Schwartz, 2001, p. 395). The term collective memory refers to the dissemination of beliefs, feelings, moral values, and knowledge regarding the past. There are a variety of generic terms in the field of memory research such as social and public memory, or "collective remembrance" (Winter and Sivan, 1999; White, 2006) which deal with theories elucidating the context of memory construction for groups, societies or nations[2]. Halbwachs ([1925] 1992), who coined the term "collective memory", asserted that individuals are incapable of remembering in a coherent successive manner outside the connections and constraints of their group and it is thus society that determines and fashions their memories. Halbwachs calls collective memory "what one has agreed to call the past'' (p. 131).

This "one" is not personal or autobiographical, nor is it based on historians or the state: Collective memory is the body of memory and remembering processes that are built and maintained by a social community, sensitive, importantly, to the context of the present. Unlike individual memory, collective memory comprises recollections of the past that are determined and produced by groups. Individuals are restricted in their ability to remember in a coherent successive manner beyond the connections and constraints of their group and thus society plays a large role in determining and shaping their memories.

If it is impossible to fix the concept with a single name, it is also practically impossible to define it with a single conceptual notion agreed across a wide range of disciplines (Olick & Robbins, 1998). Used with various degrees of sophistication, the notion of memory, more practiced than theorized, has been used to denote very different things, which nonetheless share a topical common concept: how people construct a sense of their past (Confino, 1997). Collective memory thereby presumes activities of sharing, discussion, negotiation, and often contestation (Zelizer, 1995).

This obsession with the past (Huyssen, 2003) in general might be explained by the fact that memory is perceived to be "the foundation of self and society" (Casey, 1987) and as such, it is a major theme in contemporary life, a key to personal, social, and cultural identities (Kenny, 1999). Memory, both individual and collective, is constructed and reconstructed by the dialectics of remembering and forgetting, shaped by semantic and interpretative frames and subject to a panoply of distortions (Climo & Cattell, 2002).

---

[2] The fact that the unit of analysis, methodology and theoretical assumption vary greatly in the memory research, and that social memory studies is non-paradigmatic, transdisciplinary and centerless enterprise (Olick and Robbins, 1998) did not reduce zest and enthusiasm, on the contrary, the memory study field is constantly in the expansion. On issues of the methodological problems in memory studies see examples in Confino (1997), Kansteiner (2002), and Radstone (2008).

Throughout history, technological disruptions in our communication tools such as the invention of mass print technologies, have considerably changed the way we store and interact with our collective memories from the time before the adoption of these technologies. Digital technologies, and in particular the Internet and Web-based information and communication technologies are no exceptions and they too dramatically changed the way our collective memories are formed and shaped during the digital age era, as well as the way we study collective memory at the moment.

The intersection between the digital era and collective memory produced and continues to produce numerous, both expected and unexpected outcomes, which make one ask at least three important questions: 1) what are the processes that led and gave rise to a new form of remembrance, or Digitalized Collective Memory?; 2) how the digital era is shaping and being shaped by the already existing contents of memory and memorialization practices?; 3) and how digital technologies allow us to trace and study collective memory?

In this Chapter, we provide an overview of the areas of research addressing these three questions. Nevertheless, first, we need to take a step back and review collective memory in social studies before *Digital.*

## Historical-sociological trajectories in memory studies

Le Goff (1992) identifies five distinct periods in the history of memory, from oral memory of prehistory through the advent of writing systems, the invention of the printing press together with archives, libraries, and museums, to finally electronic recording and transmitting information in the twentieth century. While the field of memory studies is rapidly growing and expanding our understanding of the importance of collective memories to our identities, meaning-making processes, and the very sense of security, much of this effort was directed to understanding the interplay between nation-states and the collectives that are governed by those states. This is not without a reason – the rise of nation-states since the late 18[th] century (Malešević, 2013) coincides with its ability to inflict an ideological blueprint, that is directly connected with the rise of organizational power. The rise of the modern state is closely linked with the project of mass literacy.

Gellner (1983) sees that homogeneity required by the modern state was only possible through a common national literacy. State formation required advanced fiscal administration, courts, other legal institutions, regional administration, and the financial infrastructure. The steady rise of the administrative power of the state is linked to the historical process of bureaucracy, infrastructure, urbanization, transport, centralization, and fiscal capacity, all of which were embedded in promoting mass literacy via national educational systems (Malešević 2011; 2012).

Gellner (1983) rightly pointed out that modern industry requires a modern, literate, technologically equipped population and the nation-state is the only agency capable of providing such a workforce, through its support for a mass, public, compulsory, and standardized education system.

The development of common educational systems, industrial and transportation infrastructures, literacy, conscription, etc, all of which resulted in ongoing attempts to define the boundaries of a homogenized nation. Central to this endeavor was the creation of state-sponsored memorialization projects, such as museums, commemorations, national calendars, history textbooks, monuments, etc. Nation-states always attempt to implement the meta-narrative as a basis for national memory in society, creating as part of their national identity "the state-sponsored memory of a national past" or "national memory" (Young, 1993). This refers to dominant or hegemonic narratives which underpin and help organize remembrance and commemorations at the level of the nation-state (Ashplant et. al, 2000). These are essentially memories that serve the state's need for control, unity, legitimization, homogenization, discipline, etc. The modern state is one of the most important agents of memory since it has the power to formalize, codify and objectify systems through which it monopolizes or seeks to monopolize not only the legitimate physical force but also legitimate symbolic force. This includes the power to name, identify, categorize and state who is who and what is what (Brubaker & Cooper, 2000).

In other words, nation-states are assumed as their 'natural place' to tailor a usable past for the sake of advancing certain goals in the present. The First and Second World Wars (and many other wars around the globe) showed the centrality of nation-states in commemorative events and memorialization processes, where those state-sponsored memory projects were almost exclusively shaped and guarded by the state, and in contrast, external attempts to shape national collective memory were understood as illegitimate interfering into national matters. Yet globalization gradually brought about several significant changes in a nationally dominant memorialization landscape.

While the post-WWII period was marked by the rapid renewals of the nation-states' powers, and hence, legitimization of a nationally bound narrative of the past events, yet at a much slower pace, human rights doctrine started challenging the monopoly of the state-sponsored memory projects. It was the foundations and institutionalization of human rights activism that put forward a much wider agenda that gradually enabled the spread of the memorialization agenda. Through networks of human rights activism, a widely-diffused human rights infrastructure became increasingly embedded in a set of regional, international and global linkages.

During the period from 1947 to the mid-1980s, the Universal Declaration on Human Rights served as the foundation for the creation of an entire framework of international and transnational human rights discourses, within which the most important human rights instruments, non-government organizations, and international publications were established, developed, and grew in power and influence (Goodale, 2006). The so-called "third wave of democratization", since the mid-seventies, has brought an explosion of previously-suppressed collective memories and adjoining dilemmas of how to address past wrongdoings (Huntington, 1991).

Since the 1980s, the human rights vision of memorialization, as the process of remembering the wrongs of the past and honoring the victims, has grown together with the prevalent idea that public and official recognition of crimes is "essential for preventing further violence" (Hazan, 2010, p. 5). One of the defining features of the international human rights movement has become this new concern for the suffering of specific others in distant lands – an agenda that, to some extent, displaces those earlier, very nation-specific struggles, even in the same places (Moyn, 2012). To this end, certain necessary historical and social conditions had to take place to bring about the rise of the moral state of compassion, defined by Sznaider (1998) as an active moral demand to address the suffering of others.

The moral demand to act to lessen the suffering of others, across spatial and temporal dimensions, became possible only in the intersection between 'humanitarianism' and the emergence of liberal society, with its distinctive features of capitalism (the market) and democracy –civic equality and citizenship (Sznaider, 1998, p. 118). On the one hand, through democratization and the lessening of profoundly categorical and corporate social distinctions, compassion becomes more extensive. On the other hand, the emergence of the market society, unintentionally, through widening the scope of exchange, also extended the public scope of compassion (Sznaider, 1998, p. 119). Further, through memories of human rights abuses and their institutionalization in international conventions, cruelty became understood as the infliction of unwarranted suffering, and compassion, and the public response to this evil – transformed into an organized campaign to lessen the suffering of strangers (Sznaider, 2015).

The rapid growth of memorialization across the globe during the 1980s and 1990s, and the obsession that shifted from commemorating victories to commemorating massive past human rights abuses, might be explained by the fact that memorialization became a crucial representation of the identity politics struggle (Winter, 2001). In particular, since the 1980s and more so the 1990s, when identity politics was a feature of the recovery of witnesses after the Soviet empire collapsed in 1989 (Winter, 2001), it became clear to all parties involved in the process of memory construction that memory is not a guaranteed right but a privilege. The 1990s were filled with illusions generated by the collapse of communist regimes, the retreat of social democracy in Europe, and the abandonment of socialist ideals in post-colonial Asia and Africa. The ethical vacuum was filled by human rights, which were entrusted, as Moyn (2012, 9) wrote, with "the grand political mission of providing a global framework for the achievement of freedom, identity, and prosperity". The increasing importance of memory also has to do both with developments in information technology and the inclusion of post-traumatic stress disorder (PTSD) in 1980 as a recognized medical diagnostic classification. Once accepted as a syndrome, PTSD validated not only rights to pensions, medical care, and public sympathy but also the commemoration of traumatic memories (Young, 1995).

The extension of the scope of compassion, together with the institutionalization of the human rights in the world polity, resulted in the formation of 'cosmopolitan memories', in particular concerning the Holocaust memory, which has become globally understood as "a unit of measurement" (Levy & Sznaider, 2002) in relation to the human rights regime. Particularly for the "third wave" democratic transitions in Eastern Europe and Latin America, which helped shape the core paradigms and normative assumptions of the field, the desired endpoint of the transition in question typically resembled a Western liberal market democracy.

Transitional justice is reflective of and contributes to, cosmopolitan imperatives as it revolves around judicial procedures and memory practices addressing legacies of human rights abuses. Facilitating transitions from authoritarian regimes to stable democratic governance has come to involve some degree of recognition of the 'other': in the context of international legitimacy, cosmopolitan imperatives command a narrative that acknowledges past injustices (Levy & Sznaider, 2010). While cosmopolitan memory provided a paradigm for understanding the global spread and the usage of the Holocaust memory trope, others argued that the Holocaust memory is utilized as a screen memory that often disables a discussion and memorialization of other atrocities. Having implications in a relatively globalized world, across many nation-states, the Holocaust is often used as a 'screen memory' to displace, repress, or 'screen' other, perhaps more traumatic, local events and histories.

The concept of screen memory serves to draw attention to the complexities of social memory, as it simultaneously produces and interrogates knowledge about the past as a way to both conceptualize and trouble contemporary notions of social memory. A screen-memory is a Freudian notion that addresses a memory of something that is unconsciously used to repress the recollection of an associated but distressing event. The concept of screen memory is seen as a bracketing mechanism that draws attention to the complexities of social memory, as it simultaneously produces and interrogates knowledge about the past as a way to both conceptualize and trouble contemporary notions of social memory (David, 2013).

This was further reflected in a new theoretical concept 'multidirectional memory' coined by Michael Rothberg (2009). Rothberg (2009, p.16) has, however, harshly critiqued this approach of "competitive memory - a zero-sum struggle over scarce resources" preferring to "consider memory as multidirectional: as subject to ongoing negotiation, cross-referencing, and borrowing". He argued that claims that remembering one thing must come at the cost of another are historically problematic, as well as politically and ethically unproductive. Instead, according to Michael Rothberg, Holocaust consciousness serves as a platform for articulating issues of national interest and thus is activated as a screen memory that does not simply compete with other pasts but also provides a greater level of comfort than that which confrontation with more local problems could allow. Rothberg claims that the Holocaust memory is ultimately interconnected with slavery, colonial domination, and forms of genocide across the globe, thus it necessarily simultaneously furthers several discourses relevant to the given national context.

Erll (2011) pushed for a new analysis on 'traveling memory' showing that the effects of globalization and the technological and media advancement moved localized memories from 'memory in culture' to 'cultures of memory', arguing that only by understanding how trans-local mnemonic forms and practices are translated and integrated into local repertoires, we can grasp

"the ways that people make sense of these experiences" (Erll, 2011, p. 5). We should also pay close attention, Erll suggests, to the various ways in which traveling memory is localized, contexts are not sufficiently described as 'another culture' but must be reconstructed as complex constellations of intersecting group allegiances, mnemonic practices, and knowledge systems.

A more comprehensive overview of the ideological blueprint in which globalization, but in particular human rights and neoliberalism impacted current memorialization processes around the globe was offered by Lea David (2020). She demonstrates the emergence of a new global phenomenon coined 'moral remembrance'. Moral remembrance refers to the human rights memorialization agenda that prescribes standards for a "proper way of remembrance" with which states are expected to comply when dealing with legacies of mass human rights abuses. It refers to the generative process of standardization of memory, that developed and adopted an isomorphic-like set of norms, based on normative world-views of human rights that, through the human rights infrastructures at the global level, promote "facing the past", "duty to remember", and "justice for victims" as its pillars.

Moral remembrance points to a gradual, accumulative development from "duty to remember" as an awareness-oriented approach to a contested past, to the policy-oriented 'proper memorialization' standards that are understood and promoted as an insurance policy against the repetition of massive human rights abuses (David, 2017). It is grounded in the presumption that "proper memorialization" is essential for "healing" societies with a difficult past and moving beyond trauma and violence. Moral remembrance points to the current preference, worldwide, for memory standardization, institutional homogenization, and norm imitation. It provides a technocratic-like set of policies and a tool kit of practices that aims to advance a human rights vision of memorialization processes to promote democratic, human rights values across the globe.

From the outset of the emergence of the human rights memorialization agenda, the main authors of human rights discourses, including the UN, western states, international NGOs, and senior western academics, constructed a three-dimensional prism of victims, perpetrators, and bystanders (Matua, 2001), which further echoed in the human rights memorialization agenda. Roughly over the past four decades, moral remembrance has adopted three main principles: 1) the necessity to collectively face a troubled past; 2) a collective duty to remember human rights abuses; and 3) a victim-centered approach that puts victims at the heart of memorialization efforts. Though all three of these principles have very different sociological-historical trajectories and are rooted in distinct ethical and philosophical ideals (David, 2020), they merged and became pillars of the human rights memorialization agenda. The emergence of moral remembrance, and its extensive promotion via human rights bodies and advocates, has shaped our current understanding of how the 'proper' way to remember past atrocities and massive human rights abuses should look. To impact memorialization processes on the ground, different organizations and institutions - through the work of knowledge-based expert groups - gradually developed standardized policy recommendations and briefings.

Moral remembrance, as a new global phenomenon, is meant to force states to face and become accountable for past human rights abuses. However, David (2020) points clearly to the cleavage it creates between national and supranational levels of memorialization. This is because nation-states see their 'natural' (and exclusive) right in promoting memorialization agendas as means to homogenize distant people. Consequently, the human rights memorialization agenda is always filtered through the needs of the nation-state, hence it becomes localized, abused, and distorted.

This wave of standardization in collective memory started to accelerate at the dawn of the current century whence digital technologies provided a place for global common memory. For example, Wikipedia was launched in 2001 as a "sum of all human knowledge", without qualifying which humans and what knowledge, which implicitly assumes human knowledge is a globally shared entity. The notion of globally shared knowledge is adjacent to the notion of global collective memory. Whilst it seemed that technological advancement was in sync with the societal changes we were going through, the interaction between those technologies and collective memory came in different shapes and forms, Wikipedia edit wars across national boundaries being one of the many examples of which (Sumi et al., 2011a). In the remainder of this Chapter, we will focus on different ways digital technologies and collective memories affected each other in the first fifths of the 21st century.

# Digitalized Collective Memories

"What if a girl in the Holocaust had Instagram?" asked the trailer of the fake Anna Frank's account, created by Mati Kochavi, an Israeli hi-tech billionaire who is from a family of Holocaust victims and survivors, and his daughter, Maya, and tagged as @eva.stiries.[3] The brief film shows simulated cellphone footage of her fictionalized life, from dancing with friends and a birthday with her grandparents to Nazi troops marching through the streets of Budapest.[4] The idea was to show a Jewish girl's life as if documented on social media to teach young people about the WWII genocide.

The project, a high-budget visual depiction of the diary of Eva Heyman – a 13-year-old Hungarian who chronicled the 1944 German invasion of Hungary, sparked endless debates, not only in Israel but around the globe. Is it morally justifiable to create simulation-like characters for the purpose of remembrance? Is it offensive to victims and survivors? Is it manipulative? However, those debates also raised an additional set of questions. Namely, what is the role of technologies in the processes of memorialization?

To start, we need to untangle two different processes. First, hand in hand with technological advancement, archives, private, state-owned, and museum collections and various forms of data, started gradually to be digitalized. This process was mainly meant to preserve the documentation and artifacts in the long run. In this way, the idea was future-oriented, to preserve memory and the imagery of the artifacts from possible destruction (Rumsey, 2016). Digitalization of archives and collections further enabled wider dissemination of information and was focused on knowledge production.

---

[3] https://www.instagram.com/eva.stories/?hl=en
[4] https://www.timesofisrael.com/instagram-story-of-young-holocaust-victim-aims-at-new-generation/

However, this process gradually led to changes in the memorialization landscape. In a way it brought about the democratization of memory – everyone in a position of basic technological devices and an internet connection could produce content and promote it on the web. The widespread adoption of the World Wide Web in the years following this work and its worldwide impact means the World Wide Web has surely cemented itself as the defining technology of this fifth phase in the history of memory (perhaps even deserving of a whole new phase) (Novick, 2000, pp. 267-268) calls into question whether the "very inorganic societies of the late twentieth century (fragmented rather than homogeneous, rapidly changing rather than stable, the principal modes of communication electronic rather than face to face)" are even capable of forming communities of collective memory. Yet, it is apparent with the development of the Web, whilst spatio-temporal coherence of individuals is less of an assortative force, the Web's affordances have allowed, even encouraged, new types of community to form - to the point that concerns around filter bubbles and echo chambers are increasingly prominent in both academic and media discourse (Flaxman et al., 2016).

The availability of social platforms altered the mnemonic communities and memory agents that were traditionally limited to the political elite and various well-connected victim groups and veteran organizations. All of the sudden, the WWW managed to connect people around the globe and create steady or sporadic communities of memory activism. Although memory activism, defined as the strategic commemoration of contested pasts outside state channels to influence public debate and political discourse (Gutman, 2017), became prevalent with the rise of the human rights memorialization agenda, the digital age made it even more accessible. People across different places forged mnemonic communities for the sake of promoting a certain memorialization agenda.

Although nascent, the fast-growing field of digital memory is best seen in the recent publications that address various aspects of this phenomenon. Topics such as digital memory in media (Hoskins, 2018) online memory activism (Fridman and Ristic 2020; Fridman, 2022), hashtag memory activism such as digitally mediated connective action (Bennett and Segerberg, 2012; Fridman, 2022; Guobin, 2016), selfie culture and memorialization (Steir-Livny, 2021), and many other related issues.

The fact that we are witnessing the emergence of a new field is evident in a recent wave of conferences, webinars, and newly instigated platforms[5] to embrace the research of digital memory cultures. Whereas the Covid-19 pandemic created a necessity to embrace online commemorations, both at the bottom-up and the top-down level, the collective memory in the digital age marks new constellations of the promotion of memory both in terms of content and in terms of agency. The ramifications of those processes are far-reaching and yet to be unveiled.

---

[5] Such as Digital MSA, an online platform for the research of digital memory at the memory Studies Association.

# Digital Collective Memories

In addition to the digitalized (version of existing) collective memory, the digital sphere has provided new affordances to how we construct and revisit collective memory, most prominently through the World Wide Web. One may recall Nora's (1989) concept of memory places, that certain objects or events can have special significance in the memory of certain groups. These "lieux de memoire" or "sites of memory" are where "memory crystallizes and secretes itself" (p. 7) and constantly changes as they gain new meaning and connections in the context of the present. Given its pervasiveness, it is not surprising that groups have adopted parts of the Web as digital lieux de memoire. However, beyond acting as a simple storage place for independently developed collective memories, the social Web has also enabled new mechanisms to facilitate the communication, (re-)negotiation, and remembrance of collective memory. The respective lieux de memoire of the Web reshapes the very collective memory they host.

Considering Jan Assmann's distinct concepts of communicative and cultural memory (Assmann and Czaplicka, 1995) illuminates how digital technology can redefine collective memory. To Assmann, cultural memory is composed of Nora's memory spaces - where events long ago may be recalled by a social group, be it through monuments, heirlooms, museums, or archives. In contrast, communicative memory "lives in everyday interaction" (Assman, 2008, p. 111), and does not require the infrastructure of cultural memory. This means, however, that it covers a relatively short time frame. Assmann turns to Vansina's work on oral history to show that communicative memory's temporal horizon spans only 80 to 100 years. The transitional period between the formalized, remote, cultural memory and more fluid, recent communicative memory, where relatively little is remembered, is identified as Vansina's concept of the "floating gap" (Vansina, 1985, p. 23). With the World Wide Web, the everyday communication and longer-term memory spaces may strongly intersect. As such, distinctions between communicative and cultural memory begin to blur.

The paradigmatic example of a site of digital collective memory is Wikipedia (or perhaps more precisely, the different language Wikipedias). Wikipedia is one of the most popular websites on the Internet, providing encyclopedic information, freely editable by anyone, across over 300 languages to hundreds of millions of unique monthly visitors. The processes of memorialization and remembrance of past and current events are constantly playing out on Wikipedia. "Fixing the floating gap: The online encyclopedia Wikipedia as a global memory place" (Pentzold, 2009) forms the basis for much of the following scholarship on collective memory on Wikipedia. Pentzold applies the work of Halbwachs on collective memory, Nora's memory places, Assmann's distinction between communicative and cultural memory, and Vansina's floating gap model to argue that Wikipedia "is a global memory place where locally disconnected participants can express and debate divergent points of view and that this leads to the formation and ratification of shared knowledge that constitutes collective memory" (p. 263).

Pentzold creates a framework for understanding the memory function of Wikipedia, arguing that Wikipedia is akin to a modified form of Nora's memory place, through the constant editing and discussion of articles "Wikipedia is not [just] a symbolic place of remembrance but a place where memorable events are negotiated'' (p. 264). He goes on to argue that "the production of articles and the parallel discussions on the associated talk pages can be viewed as the dynamic transition, the 'floating gap', between fluid communicative and static collective memory were forms of objectified culture (e.g. texts, images) are crystallized'' (p. 264).

Rosenzweig (2006) explores the prospect of Wikipedia as an 'open-source' history and how Wikipedia does not conform to traditional historical recording practices. Rosenzweig notes the "possessive individualism'' of historical scholarship, at odds with the practically anonymous volunteer editors of Wikipedia, as well as how in contrast to the gatekeeping practices of academia, there is a "denigration of expertise" – practically anyone can edit any article.[6] The author even observes the relatively disproportionate focus on current events and the opportunities for review and reframing -- "Wikipedia offers a first draft of history, but unlike journalism's draft, that history is subject to continuous revision". Rosenzweig's insights are very impactful in the early social scholarship on Wikipedia, yet he does not make the final step in linking the social knowledge creation and curation on Wikipedia to the established theory on collective memory.

Luyt (2016) has applied Pentzold's approach in a study of the Vietnam War, analyzing how different views of the war come to be contested and represented on Wikipedia as part of a collective memory-building process. In "Collective memory building in Wikipedia : The case of North African uprisings'', Ferron and Massa (2011) have also examined the process of collective memory-building on Wikipedia by profiling the collaboration and conflict of the editor network documenting the 2011 Egyptian Revolution in several Wikipedia languages. The details of editor collaboration and conflict across cultures following traumatic events are studied, from an initially disorganized and unstable community towards a more common narrative as it cements its place in our collective memory.

A more quantitative approach to the "Dynamics of conflict on Wikipedia'' is undertaken by (Yasseri, et al., 2012), who, using a previously developed measure for the controversiality of articles, observe the correspondence between conflict on Wikipedia and bursty activity, together with memory effects, as well as three distinct patterns of conflict evolution; consensus, sequence of temporary consensuses, and never-ending wars. Shi et al. (2019) find that in certain cases this conflict process can yield better Wikipedia content, with ideologically diverse teams of editors producing higher quality articles, especially in the realm of politics. The authors attribute this to "longer, more constructive, competitive and substantively focused but linguistically diverse debates'' on article talk pages[7] together with greater adherence to Wikipedia policies. There is a distinct variation in the collective memory-building process based on the topic, its significance, and the makeup of editors.

---

[6] Especially true at the time of writing, however limited rules are now in place for some contentious or important articles.

[7] An associated page for each Wikipedia article where editors may discuss what revisions to make.

## Digital Collective Memory across cultures

Whilst there are legitimate concerns about the diversity and representativeness of its editors, Wikipedia, across all of its articles and languages, is not a monolith. The most prominent illustration of this is in the variation between different language Wikipedias and even the main and *simple* English Wikipedia editions (Yasseri, Kornai & Kertész, 2012). Samoilenko et al. (2017) explore the variation of detail devoted to national histories across Wikipedia languages by measuring the incidence of years mentioned in national history articles, identifying recency and Eurocentric biases, cultural similarities between particular countries and languages, as well as variations in consensus across different language Wikipedias towards certain countries' national histories. Aragon et al. (2012) have explored how networks of famous people are recorded by different language Wikipedias. Whilst global social network measures are similar and common structures are present across all languages, the social networks in Wikipedias in languages from geographically or linguistically close communities are indeed found to be more similar.

The various language Wikipedias broadly reflect the respective cultures and communities that speak each language. But how accurate and clearly defined is this partition? Vrandečić (2021) argues that firstly, language does not align with the culture. There is a single English Wikipedia covering millions of different English speakers across the UK, USA, Australia, South Africa, India, etc., and a single Portuguese Wikipedia for speakers from Portugal and Brazil, as well as many more examples. Thus the splitting of knowledge along these lines is not an a priori rationalization for how culture and memory should be organized, but a "historical decision, driven by convenience".

Nevertheless, the aforementioned examples of national histories and famous individuals do show that language is at least partly useful in distinguishing collective memory between cultures more broadly. Vrandečić (2021) also notes the boundary setting in place by the establishment of separate Croatian, Serbian, Bosnian, and Serbocroatian language Wikipedias. Standard versions of these languages (as used on Wikipedia) are often more linguistically similar to each other than the various dialects within a particular language. Yet political and national identity have motivated the establishment of these separate lieux de mémoire as part of a continuation of wars by other means.

Aside from the complex political histories of these countries. What might have exacerbated the divisions between these Wikipedias is the emergence of a far-right bias on the Croatian Wikipedia. Media attention has been drawn to the website's historical revisionism, promotion of a fascist worldview, and whitewashing of historical and more recent crimes by far-right public figures. This came about through a gradual takeover beginning in 2009 by a small number of conservative administrators, yet it was only in 2021 that the Wikimedia Foundation took decisive action (despite prominent media coverage at least as far back as 2013). In March 2021 some editors deemed responsible for the offending content had their administrative privileges revoked and in June 2021 the Croatian Wikipedia Disinformation Assessment[8] was published. The Croatian encyclopedia has been in recovery since, attempting to change the content, as well as its public perception. This case is a rare tragedy in Wikipedia's mission of providing openly accessible knowledge to the world. This also acts as a cautionary tale for those quick to draw connections between repositories of digital collective memory and who it purports to represent.

---

[8] https://therecord.media/wikimedia-bans-admin-of-wikipedia-croatia-for-pushing-radical-right-agendas/

The abhorrent far-right content on Wikipedia may or may not be representative of the national collective memory or identity in Croatia. However, this does not preclude this content from being a representation of collective memory for groups of individuals, i.e., the group of editors who contributed it together with their supporters. Veracity, or neutrality, is not a condition for the formation of collective memory. The Internet has even facilitated the formation of communities of disconnected individuals based on extreme content such as through white nationalist forums like Stormfront (Bowman-Grieve, 2009), or conspiracy communities like those built around the QAnon phenomenon (Zuckerman, 2019). That being said, there are many more positive examples of communities building digital collective memory on Wikipedia. One example is the Black Lives Matter movement. Twyman et al. (2017) have studied the editing practices of this community, finding intensified documentation of events on Wikipedia, collaborative migration of editors to related articles, and dynamic re-appraisal of past events complementing current events.

Whilst content and editors do tend to align with language, there are also many multilingual editors responsible for the exchange of information between Wikipedias. Multilingual Wikipedia work has shown that these editors tend to be more productive and produce higher quality content than single language editors (Kim et al., 2016). Projects such as Wikidata and Abstract Wikipedia are also seeking to help unify some content across Wikipedia languages. At a more crude level, several smaller Wikipedia languages often generate new articles by automatic machine translating the equivalent article from a larger Wikipedia (typically English), before human editors go on to modify the text and broader content. This would all suggest that in certain areas there may be a larger overlap between Wikipedia language editions than one might initially expect. Yasseri et al. (2014) work on comparing contested topics in different Wikipedia language editions shows similarities and differences between the most controversial topics across languages.

Language is a straightforward and useful way to think about how collective memory maps to cultures and communities, but it is ultimately not the defining feature of which community digital collective memory on Wikipedia represents. When presented with information on Wikipedia (and often the rest of the Web), one must undertake additional work to understand whose collective memory is being represented. This is somewhat of a reversal of expectations and convention of more traditional collective memory studies. Typically, one would pre-identify a community, then set about collecting information on their collective memory through interviews, written output, art, etc. When working with Wikipedia, one frequently starts from a (relatively) centralized repository of collective memory, and must later make inferences as to which community(s) is represented. This task is made significantly more difficult by the fact that similar to much of the Web, individuals use anonymous or pseudonymous accounts to contribute to Wikipedia. Not only is this a challenge for the researcher, but it is a fundamental reworking of collective memory formation. The prominence of anonymous or pseudonymous ways to contribute to the online sphere means that gatekeeping or checks on who may contribute to a community's collective memory are less stringent. In practice, contribution and remembrance rely solely on self-identification with some aspect of the material at hand.

## Contested Digital Collective Memory

Far from being a haven of a harmonious agreement to a consistent editing narrative, Wikipedia is a site of regular disagreements between editors, even a battleground for "edit wars". For the most part, this contestation is relatively minor, and the "wisdom of polarized crowds" (Shi et a., 2019) frequently elicits better quality articles that can integrate the nuances of each position, receive moderating input from more editors, and appeals to a wider variety of sources. The contestation of content (as well as revisiting and renegotiating it) is a core part of collective memory building both in the memory space and in the minds of the collective who build and access the content.

In terms of conflict online, Wikipedia certainly fares much better than the rest of social media (Yasseri & Menczer, 2021) where concerns are rife over polarization, echo chambers, and filter bubbles (that is not to say this content cannot be an example of collective memory). There are different patterns of contestation of digital collective memory that might arise, Yasseri et al. (2012) found three distinct patterns of conflict evolution; consensus, sequence of temporary consensuses, and never-ending edit wars — where disagreements between different camps are too large and result in editors constantly reverting each other's changes to an article.

Wikipedia does not, and cannot, represent the collective memory of all groups, despite the common belief and the over-idealized goal that it should (Iñiguez et al., 2014). Very occasionally a delicate, unstable consensus may emerge, but this may rely on interventions from more senior editors or administrators (still all volunteers unaffiliated with Wikimedia). Here is where we start to see a departure from a purely democratic construction of collective memory, and where appeals to systems and processes managed by higher levels of authority are required. Controversial Wikipedia pages where edit wars or vandalism are out of control can be put under various temporary or indefinite protection measures by administrators (Sumi et el., 2011b). These measures range in severity from only allowing logged-in users to edit ("Semi-protection"), to only allowing administrators to edit ("Full protection" - effectively a moratorium on editing until a dispute is resolved on the article talk page). In practice, full protection is implemented sparingly and temporarily, however, there are many pages under indefinite "Extended-confirmed protection" (e.g. Volodymyr Zelenskyy, State of Palestine, COVID-19 pandemic). A user account must be at least 30 days old and with at least 500 edits to its name to edit an article under this level of protection.

These measures may be necessary to ensure the smooth running of the encyclopedia, and Wikipedia is rightly praised for the overall effective management of content on the site, but they ultimately represent a failure in the free formation of collective memory. The irresolvable differences between groups of editors are incompatible with Wikipedia's single account for each subject. Certain groups may choose to leave the site, as well as establish their own alternative Wikis that better reflect their point of view for their own site of collective memory, such as Conservapedia[9] (conservative Christian fundamentalist, often conspiracist) or the New World Encyclopedia[10] (formed by the Unification Church, a new religious movement in support of Korean reunification). In these cases, the aforementioned concerns on determining community identity are less problematic.

## The Place of the State in Digital Collective Memory

The previous examples of alt-Wikis are broadly grassroots and community-driven, but national governments have also attempted to control and shape the online encyclopedic record and further pervert collective memory construction. Wikipedia has been subject to various bans or controls in countries around the world, frequently during times of national upheaval or controversy. Examples include a blanket ban on all Wikipedia languages in the wake of a military coup in Myanmar[11] or the ban of Wikipedia in Turkey[12]. The single language ban is particularly interesting—the Syrian government opting to restrict the more culturally representative Arabic Wikipedia which would be more accessible and relatable to the majority Arabic-speaking population, rather than e.g. minority language Kurdish Wikipedia, or the larger English and French Wikipedias which may be understood by more educated, urban Syrians. Russia has also had a tense relationship with Wikipedia, with various attempts and threats to Wikimedia and administrators to remove or rewrite individual articles. This has most recently resulted in the arrest of Mark Bernstein in Belarus, for violating Russia's "fake news law" by editing the Wikipedia article for the 2022 Russian invasion of Ukraine[13]. These attempts are not limited to authoritarian regimes. In 2013, the French interior intelligence agency pushed for an article about a military radio facility to be deleted over national security concerns, despite much of the information being publicly available elsewhere.[14] These measures, in restricting access to information by editors and regular users, damage processes of collective memory formation and remembrance.

Of particular interest is China's relationship with digital collective memory through the online encyclopedia. The Chinese Wikipedia (zh.wikipedia.org) was launched in May 2001 and at the time of writing is the 13th largest Wikipedia (number of articles) and 7th most popular (active editors).[15] Since 2004, Wikipedia has faced numerous sitewide or article-specific bans in China (e.g. on politically unfavorable articles such as on the 1989 Tiananmen Square protests), both temporary and indefinite, and is currently subject to a ban across all languages since April 2019.[16] What sets China apart is the launch of a popular Wikipedia alternative—Baidu Baike—in April 2008. Nominally, Baidu Baike claims it is an open encyclopedia (in the same vein as Wikipedia) with core values of "equality, collaboration, sharing, and freedom",[17] however, the thresholds for editing are much stricter than Wikipedia's. Baidu Baike uses a content review policy, where all edits (from users that must use a registered account) have to be approved by company-appointed administrators. Articles must also meet various guidelines and should not contain "reactionary content" that "maliciously evaluates the current national system", or content that "violates ethics and morality" or "social public morals".[18] As might be expected, the politically sensitive content on Wikipedia that attracted scrutiny from Chinese authorities is not to be found (or is very heavily sanitized) on Baidu Baike. Many articles have also been plagiarized from Wikipedia, with Baidu Baike being called the "Worst Wikipedia Copyright Violator".[19]

---

[9] https://www.conservapedia.com/Main_Page
[10] https://www.newworldencyclopedia.org/
[11] https://dig.watch/updates/wikipedia-blocked-myanmar
[12] https://www.reuters.com/article/us-turkey-wikipedia-idUSKBN1YU0S3
[13] https://www.theverge.com/2022/3/11/22973293/wikipedia-editor-russia-belarus-ukraine
[14] https://edri.org/our-work/edrigramnumber11-7french-spys-want-takedown-wikipedia-content/
[15] https://meta.wikimedia.org/wiki/List_of_Wikipedias
[16] https://www.bbc.com/news/technology-48269608

With its similarity to Wikipedia, Baidu Baike could effectively be positioned as an equivalent site of collective memory. However, the level of state interference in this encyclopedic record means this collective memory can hardly be considered 'organic'. For all its strengths, Wikipedia and related sites' positions as relatively centralized sites of digital collective memory is precarious and can be subject to the whims of internal administrators and external authorities.

# Studying Collective Memory in the Digital Age

The Internet has brought change to almost everything in our lives. It has even changed the way science is being done. Social scientists increasingly are using data that are generated as a result of our online activities to study our individual or collective behavior on a scale and with an accuracy normally only seen in the natural sciences. Sure, we are still far from having large experimental social science data sets similar to the ones produced in CERN, but at least we have digital observational data like that collected and analyzed in observational astrophysics.

Even though collective memory is a fundamental concept in sociology, there have been very few empirical studies on the subject, mostly because of a lack of systematically generated large-scale data or the expertise required to leverage the existing data systematically. Traditionally, scientists who research how the public recalls past events had to spend a lot of time and effort collecting data through interviews, archives, ethnographies, and visual elicitation.

Utilizing such digital data, and in particular traces of information-seeking behavior on online knowledge, repositories have contributed significantly to the field of collective memory studies in recent years. In addition to the construction of collective memory, the process of revisiting events in collective remembrance has also been investigated using Wikipedia page view statistics. Kanhabua et al. (2014) study the processes that trigger the act of revisiting past events by considering related events (e.g. hurricanes, and terrorist attacks) over time, across several different categories of traumatic events. Different event types are observed to have different characteristics of remembrance. Event memory triggering is identified across different categories using page view time-series correlation, 'surprise' detection, and distribution kurtosis, as well as more specific metrics such as event location, fatalities, and article semantic similarity. In "The Memory Remains: Understanding Collective Memory in the Digital Age'', García-Gavilanes et al. (2017) also consider collective remembrance for traumatic events and model the flow of viewership to past aircraft crashes in response to those in the news, based on their similarity in time, geography, topic, and linked articles. They show that on average the secondary flow of attention to remembered past events in response to current events is larger than the primary flow of attention to the current event itself. In related work, García-Gavilanes et al. (2016) quantify the half-life of the short-term collective memory to the current event to 5-8 days utilizing similar data from Wikipedia.

---

[17] https://web.archive.org/web/20160708105140/http://help.baidu.com/question?prod_en=baike&class=260

[18] https://baike.baidu.com/item/%E7%99%BE%E5%BA%A6%E7%99%BE%E7%A7%91%EF%BC%9A%E8%BF%9D%E8%A7%84%E8%A1%8C%E4%B8%BA%E5%8F%8A%E5%85%B6%E5%A4%84%E7%BD%9A/8384039

[19] https://abcnews.go.com/Technology/PCWorld/story?id=3451658

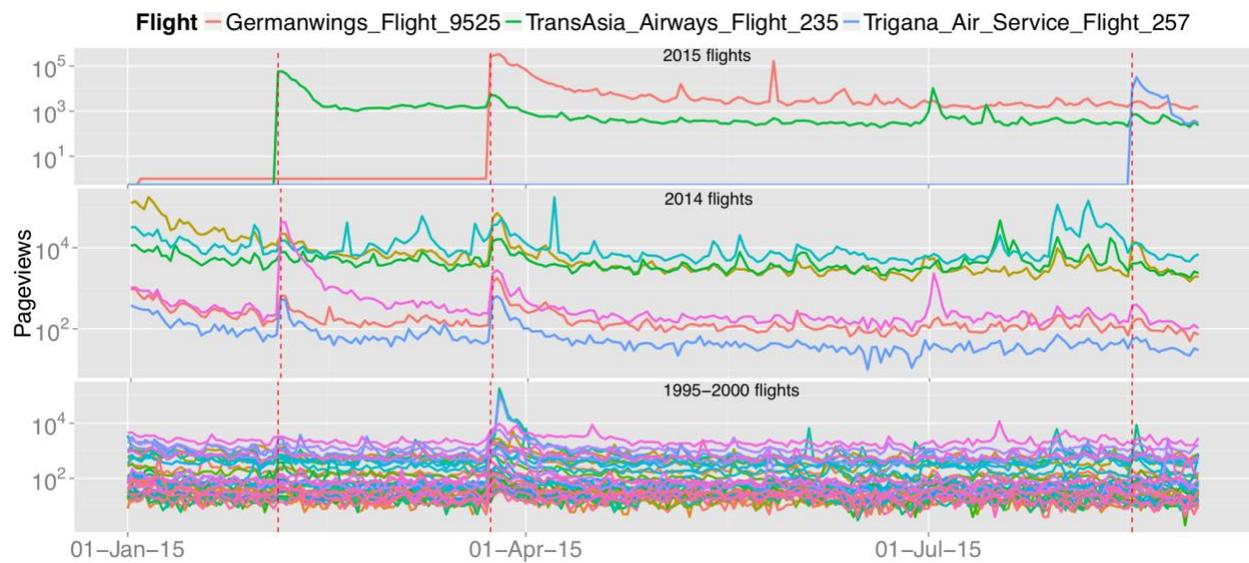

Figure 1: Page view statistics of the 1496 articles about airline disasters on Wikipedia during the first nine months of 2015. The curves show how the occurrence of a new airline disaster triggers the attention toward the past events. See García-Gavilanes at al. (2017) for more details.

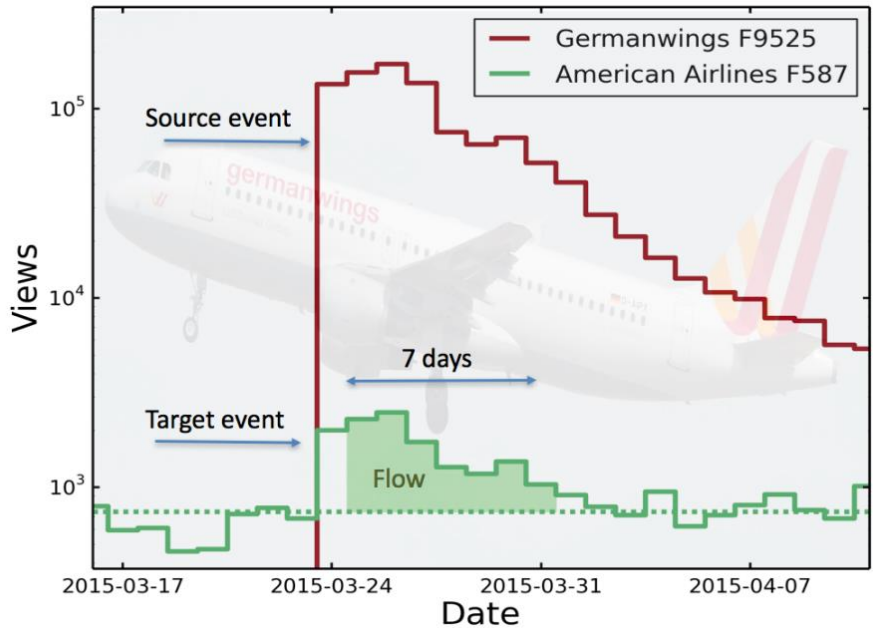

Figure 2. How the "flow of attention" from a current event to a past event is measured by García-Gavilanes et al. (2017).

Candia et al. (2019) explore longer-term, multi-platform dynamics of collective memory. The authors study cross-sectional data towards biographical Wikipedia pages, Spotify song popularity, and YouTube movie trailer plays, as well as time-series data for citations towards scientific papers and patents, with a two-step decay function they argue, corresponds to the different modes of communicative and cultural memory. Unlike Assmann's claim that this transition occurs over ~80-100 years, different critical times from around 5 to 30 years are observed for different cultural products. In addition, the authors do not take into account different dynamics that might occur around the points in time the digital platforms are introduced. It is thus not clear whether this is evidence of Assmann's floating gap, whether collective behavior in the information age has changed, or whether there are multiple modes of attention and memory as yet unstudied.

Instead of studying the attention and remembrance of defined events or cultural objects, Miz et al. (2019) put forward a graph-based dynamical pattern extraction model, inspired by Hebbian learning theory, to generate networked communities of collective memory on Wikipedia based on correlated patterns of attention and remembrance. This, in principle, may be used to detect and analyze the network of articles associated with particular patterns of attention (Gildersleve & Yasseri, 2018), however, the model is reliant on a static article network structure and does not explicitly connect to specific events, leaving inference to the researcher. These works show us the relation between various features of events and their corresponding non-trivial multiple article attention dynamics in remembrance is both significant and informative.

Whilst the framework for studying collective memory through Wikipedia is well established, the picture is incomplete. Work relating to Wikipedia and collective memory, like the approaches of sociology and psychology, either focus on the practices editors engage in when collaborating (or conflicting) in building the memory space, or the collective patterns of remembering past events in response to current events, rather than what is memorialized and how.

## Conclusion and future avenues

There remain several important areas where scholars of collective memory have failed to resolve differences or provide concrete answers. Two of these, as identified by Kansteiner (2002), are that firstly "collective memory studies have not yet sufficiently conceptualized collective memories as distinct from individual memory'' (p. 180), meaning that psychological methods are susceptible to overstating and misrepresenting the role of the individual. Social study into the way individuals' different accounts of events are brought together, contested, and negotiated to eventually form more crystallized cultural memory is required to understand the dynamics and formation of collective memory. Secondly, "collective memory studies have also not yet paid enough attention to the problem of reception'' (p. 180), the way different representations of events in records of collective memory are received by wider members of culture requires research beyond that of case studies for individual events. Building on this, we identify that whilst theoretical background and study of established collective memories is well covered by existing literature, little has been made of what makes a 'successful' collective memory. Sturken (1997, p. add1) states that "Cultural memory is a field of cultural negotiation through which different stories vie for a place in history'', and understanding this competition for attention and memory resources between different topics is a key issue. What are the short and long-term dynamics of world news events translating into coherent collective memories that mean some never capture wide attention, some attract short-term interest but fade to obscurity, some become ingrained in cultural consciousness, and some are subject to never-ending dispute? In layman's terms—how does news become history through collective actions, market forces, or active agents?

Several other questions are still to be addressed. Who has ownership, and who can access digital collective memory, as well as who should be able to? Platforms such as Wikipedia are free to all humans for now, but there is no guarantee that they remain so forever. How is the consensus over collective memory achieved on such platforms and if the achieved consensus is representative of all ideas? What will be the role of the governments, NGOs, and private stakeholders in shaping and defining collective memories in the digital world? And last but not least, what is the role of semi-automated and semi-intelligent (Tsvetkova et al., 2017) agents in forming and shaping of our collective memories?

Finally, when it comes to observational studies based on trace data such as Wikipedia page view statistics, It is important to note that we do not understand the underlying mechanisms behind these observations. The role of the media, individual memory, or the structure and categorization of articles on Wikipedia can all play a part and will be subject to future research. More traditional theories suggest that the media plays a central role in shaping our collective memory. However, a big question to ask now is whether the transition to online media and in particular social media will change this mechanism. These days, we often receive news through our Facebook friends, so can this explain why events that have not been in the news for years suddenly become so visible? Knowing the answers to these questions and understanding how collective memory is being shaped not only is interesting from a scientific perspective, but also could have applications in journalism, media development, policymaking, and even advertisement.

# References

Aragon, P., Laniado, D., Kaltenbrunner, A., & Volkovich, Y. (2012, August). Biographical social networks on Wikipedia: a cross-cultural study of links that made history. In *Proceedings of the eighth annual international symposium on Wikis and open collaboration* (pp. 1-4).

Ashplant, T.G., Dawson, G. and Roper, M. eds., 2000. *The politics of war memory and commemoration* (Vol. 7). London: Routledge.

Assmann, J. (2008). Communicative and cultural memory. In A. Erll, A. Nünning, & S. B. Young (Eds.), Cultural memory studies: An international and interdisciplinary handbook (Vol. 8, p. 109-118). de Gruyter

Assmann, J., & Czaplicka, J. (1995). Collective memory and cultural identity. *New German critique*, (65), 125-133.

Bennett, W. L., & Segerberg, A. (2012). The logic of connective action: Digital media and the personalization of contentious politics. *Information, communication & society*, *15*(5), 739-768.

Bowman-Grieve, L., 2009. Exploring "Stormfront": A virtual community of the radical right. *Studies in conflict & terrorism*, *32*(11), pp.989-1007.

Brubaker, R. and Cooper, F., 2000. Beyond" identity". *Theory and society*, *29*(1), pp.1-47.

Candia, C., Jara-Figueroa, C., Rodriguez-Sickert, C., Barabási, A. L., & Hidalgo, C. A. (2019). The universal decay of collective memory and attention. *Nature Human Behaviour*, *3*(1), 82-91.

Casey, E.S., 1987. The world of nostalgia. *Man and world*, *20*(4), pp.361-384.

Climo, J. J., & Cattell, M. G. (Eds.). (2002). *Social memory and history: Anthropological perspectives*. Rowman Altamira.

Confino, A. (1997). Collective memory and cultural history: Problems of method. *The American historical review*, *102*(5), 1386-1403.

Connerton, P. (1989). *How societies remember*. Cambridge University Press.

David, L., 2013. Holocaust Discourse as a Screen Memory: The Serbian Case. *History and Politics in the Western Balkans: Changes at the Turn of the Millenium*, pp.64-88.

David, L., 2017. Against standardization of memory. *Human Rights Quarterly*, *39*(2), pp.296-318.

David, L., 2020. Moral Remembrance and New Inequalities. *Global Perspectives*, *1*(1).

Dunbar, R.I., 2006. Brains, cognition and the evolution of culture. *Evolution and culture*, pp.169-180.

Erll, A., 2011. Travelling memory. *parallax*, *17*(4), pp.4-18.

Ferron, M., & Massa, P. (2011, October). Collective memory building in Wikipedia: the case of North African uprisings. In *Proceedings of the 7th international symposium on wikis and open collaboration* (pp. 114-123).

Flaxman, S., Goel, S., & Rao, J. M. (2016). Filter bubbles, echo chambers, and online news consumption. *Public opinion quarterly*, *80*(S1), 298-320.

Freeman, Lindsey, Nienass, Benjamin and Melamed, Lilav. 2013. "Screen memory." International Journal of Politics, Culture and Society 26: 1–7.

Fridman Orli (2022) Hashtag Memory Activism. Online Commemorations and Online Memory Activism. Magazine of the European Observatory on Memories.

Fridman, O., & Ristić, K. (2020). Online transnational memory activism and commemoration. *Agency in Transnational Memory Politics*, *4*, pp 68-92.

García-Gavilanes, R., Mollgaard, A., Tsvetkova, M., & Yasseri, T., 2017. The memory remains: Understanding collective memory in the digital age. *Science advances*, *3*(4), e1602368.

García-Gavilanes, R., Tsvetkova, M. and Yasseri, T., 2016. Dynamics and biases of online attention: the case of aircraft crashes. *Royal Society open science*, *3*(10), p.160460.

Gellner, E., 1983. Nationalism and the two forms of cohesion in complex societies. London: British Academy.

Gildersleve, P., & Yasseri, T. (2018). Inspiration, captivation, and misdirection: Emergent properties in networks of online navigation. In *International workshop on complex networks* (pp. 271-282). Springer, Cham.

Goodale M (2006) Toward a Critical Anthropology of Human Rights *Current Anthropology* 47(3):485-511.

Gutman Yifat (2017) Memory Activism: Reimagining the Past for the Future in Israel-Palestine. Vanderblit University, pp. 1-2.

Halbwachs, M., 1925. 1992. *On collective memory.*

Hazan P. (2010) J*udging War, Judging History, Behind Peace and Reconciliation*. Stanford University Press.

Hoskins, Andrew (2018). "The restless past: an introduction to digital memory and media," in Andrew Hoskins (ed) Digital Memory Studies: Media Pasts in Transition. New York: Routledge, Taylor & Francis Group.

Huntington S. (1991). The Third Wave. Norman: University of Oklahoma Press.

Huyssen, A. (2003). *Present pasts: Urban palimpsests and the politics of memory*. Stanford University Press.

Iñiguez, G., Török, J., Yasseri, T., Kaski, K., & Kertész, J. (2014). Modeling social dynamics in a collaborative environment. *EPJ Data Science*, *3*, 1-20.

Kanhabua, N., Nguyen, T. N., & Niederée, C. (2014, September). What triggers human remembering of events? A large-scale analysis of catalysts for collective memory in Wikipedia. In *IEEE/ACM Joint Conference on Digital Libraries* (pp. 341-350). IEEE.

Kansteiner, W. (2002). Finding meaning in memory: A methodological critique of collective memory studies. *History and theory*, *41*(2), 179-197.

Kenny, M. G. (1999). A place for memory: The interface between individual and collective history. *Comparative studies in society and history*, *41*(3), 420-437.

Kim, S., Park, S., Hale, S.A., Kim, S., Byun, J. and Oh, A.H., 2016. Understanding editing behaviors in multilingual Wikipedia. *PloS one*, *11*(5), p.e0155305.

Kuljić, T., 2006. Cultural Memory: Theoretical Explanations of the Use of the Past. *Beograd: Čigoja štampa*.

Le Goff, J., 1992. *History and memory*. Columbia University Press.

Levy D. and Sznaider N. (2010) Human Rights and Memory. The Pennsylvania University Press: Pennsylvania.

Levy Daniel and Sznaider Natan (2002) "Memory Unbound: The Holocaust and the Formation of Cosmopolitan Memory." *European Journal of Social Theory* 5:87-106.

Lowenthal, D., 1985. 1985: The past is a foreign country. Cambridge, Cambridge University Press.

Luyt, B. (2016). Wikipedia, collective memory, and the Vietnam war. *Journal of the Association for Information Science and Technology*, *67*(8), 1956-1961.

Malešević, S., 2011. Nationalism, war and social cohesion. *Ethnic and racial studies*, *34*(1), pp.142-161.

Malešević, S., 2013. *Nation-states and nationalisms: organization, ideology and solidarity*. Polity.

Matua, M.W., 2001. savages, saviors and victims: The Metaphor of Human Rights. *HARV. INT'L LJ*, *42*, pp.201-227.

Miz, V., Ricaud, B., Benzi, K., & Vandergheynst, P. (2019, May). Anomaly detection in the dynamics of web and social networks using associative memory. In *The World Wide Web Conference* (pp. 1290-1299).

Moyn S. (2012) *The Last Utopia: Human Rights in History.* Harvard University Press.

Nora, P. (1989). Between memory and history: Les lieux de mémoire. *representations*, *26*, 7-24.

Novick, P. (2000). The Holocaust and collective memory: the American experience (pp. 133-4). London: Bloomsbury.

Olick, J. K., & Robbins, J. (1998). Social memory studies: From "collective memory" to the historical sociology of mnemonic practices. *Annual Review of sociology*, *24*(1), 105-140.

Pentzold, C. (2009). Fixing the floating gap: The online encyclopaedia Wikipedia as a global memory place. *Memory studies*, *2*(2), 255-272.

Radstone, S. (2008). Memory studies: For and against. *Memory studies*, *1*(1), 31-39.

Rosenzweig, R. (2006). Can history be open source? Wikipedia and the future of the past. *The journal of American history*, *93*(1), 117-146.

Rothberg, M., 2009. Multidirectional memory. In *Multidirectional Memory*. Stanford University Press.

Rumsey, A. S. (2016). *When we are no more: How digital memory is shaping our future*. Bloomsbury Publishing USA.

Samoilenko, A., Lemmerich, F., Weller, K., Zens, M., & Strohmaier, M. (2017, May). Analysing timelines of national histories across Wikipedia editions: A comparative computational approach. In *Proceedings of the International AAAI Conference on Web and Social Media* (Vol. 11, No. 1, pp. 210-219).

Schwartz, B. (1982). The social context of commemoration: A study in collective memory. *Social forces*, *61*(2), 374-402.

Shi, F., Teplitskiy, M., Duede, E., & Evans, J. A. (2019). The wisdom of polarized crowds. *Nature human behaviour*, *3*(4), 329-336.

Steir-Livny, L. (2021). Traumatic past in the present: COVID-19 and Holocaust memory in Israeli media, digital media, and social media. *Media, Culture & Society*, 01634437211036997.

Sturken, M. (1997). *Tangled memories: The Vietnam War, the AIDS epidemic, and the politics of remembering*. Univ of California Press.

Sumi, R., Yasseri, T., Rung, A., Kornai, A., & Kertesz, J. (2011a). Edit Wars in Wikipedia. *2011 IEEE Third International Conference on Privacy, Security, Risk and Trust and 2011 IEEE Third*

*International Conference on Social Computing*, 724–727. doi:10.1109/PASSAT/SocialCom.2011.47

Sumi R, Yasseri T, Rung A, et al. (2011b) Characterization and prediction of Wikipedia edit wars. In: *Proceedings of the 3rd International Web Science Conference*. New York, NY: ACM Press, pp. 1–3.

Szbaider Natan (2015) Compassion, Cruelty, and Human Rights. R.E. Anderson (ed.), *World Suffering and Quality of Life*, Springer.

Sznaider Natan (1998) The Sociology of Compassion: A Study in the Sociology of Morals, *Journal for Cultural Research*, 2:1, 117-139.

Sznaider, N., 2015. Compassion, cruelty, and human rights. In *World suffering and quality of life* (pp. 55-64). Springer, Dordrecht.

Tsvetkova, M., García-Gavilanes, R., Floridi, L., & Yasseri, T. (2017). Even good bots fight: The case of Wikipedia. *PloS one*, *12*(2), e0171774.

Twyman, M., Keegan, B. C., & Shaw, A. (2017, February). Black Lives Matter in Wikipedia: Collective memory and collaboration around online social movements. In *Proceedings of the 2017 ACM conference on computer supported cooperative work and social computing* (pp. 1400-1412).

Vansina, J. M. (1985). *Oral tradition as history*. Univ of Wisconsin Press.

Vrandečić, D., 2021. Building a multilingual Wikipedia. *Communications of the ACM*, *64*(4), pp.38-41.

White, G., 2006. Epilogue: memory moments. *Ethos*, *34*(2), pp.325-341.

Winter J (2001) The Generation of Memory: Reflections on the "Memory Boom" Contemporary Historical Studies. *Canadian Military History* 10(3):57-66.

Winter, J., & Sivan, E. (1999). Setting the framework. *Studies in the Social and Cultural History of Modern Warfare*, *5*, 6-39.

Yang, G. (2016). Narrative agency in hashtag activism: The case of #BlackLivesMatter. *Media and communication*, *4*(4), 13.

Yasseri, T., & Menczer, F. (2021). Can crowdsourcing rescue the social marketplace of ideas?. *arXiv preprint arXiv:2104.13754*.

Yasseri, T., Kornai, A., & Kertész, J. (2012). A practical approach to language complexity: a Wikipedia case study. *PloS one*, *7*(11), e48386.

Yasseri, T., Spoerri, A., Graham, M. and Kertész, J., 2014. The most controversial topics in Wikipedia. *Global Wikipedia: International and cross-cultural issues in online collaboration*, *25*, pp.25-48.

Yasseri, T., Sumi, R., Rung, A., Kornai, A., & Kertész, J. (2012). Dynamics of conflicts in Wikipedia. *PloS one*, *7*(6), e38869.

Zelizer, B. (1995). Reading the past against the grain: The shape of memory studies. *Critical studies in mass communication*, *12*, 214-214.

Zerubavel, Y., 1995. *Recovered roots: Collective memory and the making of Israeli national tradition*. University of Chicago Press.

Zuckerman, E., 2019. QAnon and the emergence of the unreal.